\documentclass[aps,prapplied,reprint,superscriptaddress,floatfix]{revtex4-2}

\usepackage{amsmath,amssymb,amsfonts}
\usepackage{graphicx}
\usepackage{float}
\usepackage{xcolor}
\usepackage{xspace}

\usepackage[unicode=true,colorlinks=true,citecolor=blue,linkcolor=blue,urlcolor=blue]{hyperref}

\usepackage[normalem]{ulem}
\usepackage[version=4]{mhchem} % very useful to write formulas, like \ce{H2O}
\usepackage{siunitx}
\usepackage{silence}
\begin{document}

\title{Magnetic flux distribution, quasiparticle spectroscopy, and\\ quality factors in Nb films for superconducting qubits}

\author{Amlan Datta}
\affiliation{Ames National Laboratory, Ames, IA 50011, USA}
\affiliation{Department of Physics \& Astronomy, Iowa State University, Ames, IA 50011, USA}

\author{Bicky S. Moirangthem}
\affiliation{Ames National Laboratory, Ames, IA 50011, USA}
\affiliation{Department of Physics \& Astronomy, Iowa State University, Ames, IA 50011, USA}

\author{Kamal R. Joshi}
\affiliation{Ames National Laboratory, Ames, IA 50011, USA}

\author{Anthony P. McFadden}
\affiliation{National Institute of Standards and Technology, Boulder, Colorado 80305, USA}

\author{Florent Lecocq}
\affiliation{National Institute of Standards and Technology, Boulder, Colorado 80305, USA}

\author{Raymond W. Simmonds}
\affiliation{National Institute of Standards and Technology, Boulder, Colorado 80305, USA}

\author{Makariy A. Tanatar}
\affiliation{Ames National Laboratory, Ames, IA 50011, USA}
\affiliation{Department of Physics \& Astronomy, Iowa State University, Ames, IA 50011, USA}

\author{Matthew J. Kramer}
\affiliation{Ames National Laboratory, Ames, IA 50011, USA}

\author{Ruslan Prozorov}
\email[Corresponding author: ]{prozorov@ameslab.gov}
\affiliation{Ames National Laboratory, Ames, IA 50011, USA}
\affiliation{Department of Physics \& Astronomy, Iowa State University, Ames, IA 50011, USA}
\date{24 March 2026}

\begin{abstract}
Niobium is a practical material platform for superconducting microwave circuits; however, device-level performance can vary significantly depending on film growth and processing conditions. We compare three epitaxial Nb films grown on $c-$plane sapphire substrates under nominally identical conditions, except for the deposition temperature. To correlate internal quality factors, $Q_{\mathrm {i}}$, with material properties, we combine magneto-optical imaging of magnetic flux distribution with quasiparticle spectroscopy via measurements of the London penetration depth, $\lambda(T)$. In the low-$Q_{\mathrm i}$ film, there is a lesser ability to screen the magnetic field and an irregular temperature variation of $\lambda(T)$, implying the existence of localized in-gap states. High $Q_{\mathrm i}$ films show the opposite trend. We conclude that our measurements provide an efficient method for characterizing and optimizing superconducting films for quantum informatics applications. 
\end{abstract}

\maketitle
\section{Introduction}

Superconducting qubits and their associated microwave circuitry are limited by material-dependent dissipation \cite{Kjaergaard2020,Devoret2013,Leon2021}. For planar devices, the internal quality factor $Q_{\mathrm i}$ of coplanar resonators is a measurement of material-related microwave loss in transmon qubits. At the level of materials physics, the dominant contributions typically include dielectric loss from two-level systems (TLS) in amorphous oxides and adsorbates \cite{Martinis2005,Muller2019,Graaf2020}, non-equilibrium quasiparticles and related sub-gap states \cite{Catelani2011,Liarte_2017,Senthil24}, and vortex-related dissipation in the presence of residual magnetic fields \cite{Bothner2011,Stan2004}. These mechanisms are strongly influenced by structures on the scale of \SI{}{\nano\meter} - \SI{}{\micro\meter} on surfaces, interfaces, and grain boundaries; therefore, bulk superconducting metrics such as $T_c$ or the residual-resistivity ratio can be poor predictors of $Q_{\mathrm i}$.

Niobium is widely used in superconducting technology due to its relatively high $T_c \approx 9.2$~K and compatibility with established thin-film processing techniques. At the same time, a growing body of work indicates that native Nb oxides and sub-oxides can host TLS-active defects and other low-energy excitations that limit microwave performance \cite{Bal2024,Gorgichuk2023,Bafia2024,Kalboussi2025}. This motivates the development of diagnostic tools sensitive to vortex physics and the low-energy electrodynamic response of the superconducting condensate, both of which depend on the details of film fabrication.

In this paper, we analyze three epitaxial Nb films deposited at different substrate temperatures and previously categorized as low, intermediate, and high-$Q_{\mathrm i}$ based on single-photon resonator measurements \cite{mcfadden2025}. Our focus is to establish correlations between (i) magnetic flux entry/trapping and its mesoscale uniformity, and (ii) the temperature dependence of the London penetration depth $\lambda(T)$ and the resulting superfluid density. We used magneto-optical (MO) imaging to map the spatial distribution of magnetic induction, $B_{z}(x,y)$, and a tunnel-diode resonator (TDR) to measure $\Delta\lambda(T)$, which is essentially quasiparticle spectroscopy \citep{Prozorov2006,Joshi2023,ghimire2024quasi}.

\section{Samples and methods}

\subsection{Thin films and microwave characterization}

\begin{figure}[b]
\includegraphics[width=0.8\columnwidth]{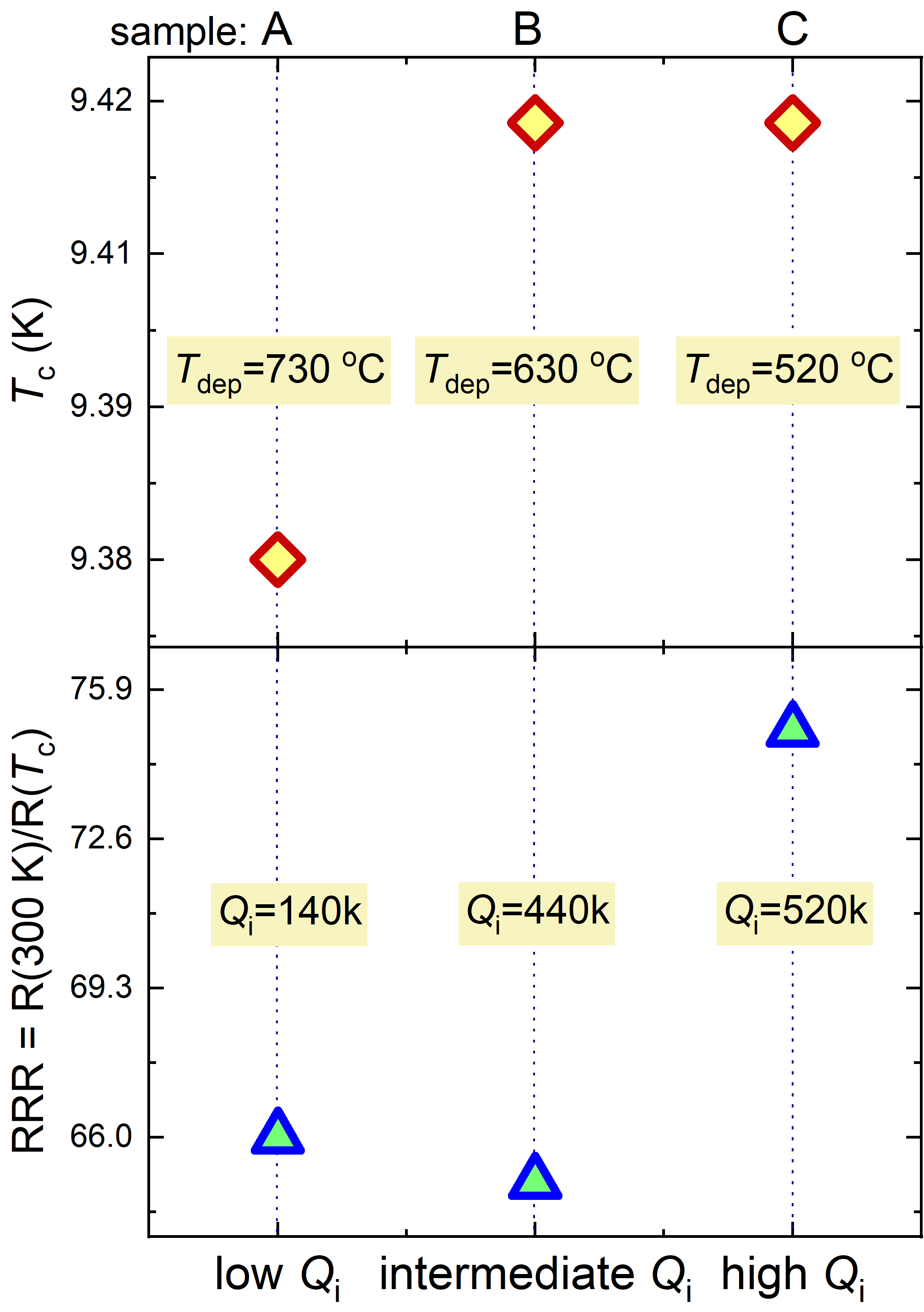} 
\caption{\label{fig1:Samples}Nb films studied in this work. Top panel shows superconducting transition temperature, $T_c$; bottom panel shows corresponding $RRR=R(300\,\text{K})/R(T_c)$. Data from Ref.~\cite{mcfadden2025}.}
\end{figure}

Thin niobium films, 100~nm thick, were grown on identical $c$-plane sapphire substrates under the same base pressure and sputtering conditions at different substrate temperatures. Following Ref.~\cite{mcfadden2025}, we refer to the three growth conditions as low-$T_{\mathrm{dep}}$, intermediate-$T_{\mathrm{dep}}$, and high-$T_{\mathrm{dep}}$, corresponding to high-, intermediate-, and low-$Q_{\mathrm i}$ resonator performance, respectively. All three films were found to be epitaxial Nb(111) having a single crystallographic orientation both in and out of plane, as determined by X-ray diffraction according to the structural characterization in Ref.~\cite{mcfadden2025}. The basic properties of the films studied are summarized in Fig.~\ref{fig1:Samples}. Note that despite almost equal temperature increments across the three sample types, the quality factor is highest in sample C, which has the lowest deposition temperature. However, there is no direct correlation between $Q_{\mathrm i}$, $T_c$ and $RRR=R(300\,\text{K})/R(T_c)$. Still, sample C has the highest $RRR$ and $T_c=9.42\,\text{K}$, which is remarkable for Nb films that usually show $T_c \leq 9.35\,\text{K}$~\citep{mcfadden2025,Tanatar2022}.

\subsection{Magneto-optical imaging of the magnetic flux distribution}

A two-dimensional (2D) distribution of the normal component of magnetic induction, $B_{z}(x,y)$, on the sample surface was visualized using a Faraday effect-active indicator film (transparent ferrimagnetic bismuth-doped iron garnet) placed on top of the superconducting film \cite{Jooss2002,Young2005,Prozorov2006a}. Linearly polarized light propagating through the indicator rotates by an angle proportional to the local value of $B_{z}$. In all images, brighter colors correspond to a higher magnitude of the local magnetic induction. Details of the technique and calibration procedures are given elsewhere \cite{Prozorov2006a,Jooss2002}. Magnetic flux was imaged after (i) zero-field cooling (ZFC) to measurement temperature, followed by application of an external perpendicular magnetic field $H_{a}$, and (ii) field cooling (FC) through $T_c$ in a fixed $H_{a}$, zeroing the field and followed by imaging at low temperature.

To compare different films, we used magneto-optical images to examine the overall intensity maps of the $B_{z}(x,y)$ distribution, which readily reveal mesoscopic inhomogeneities that are difficult to access with macroscopic magnetization measurements or nanoscale spectroscopy. It is also used to ascertain the film's ability to shield the externally applied magnetic field by measuring the average depth of flux penetration from the edge at the same temperature and magnetic field after ZFC. 

\subsection{London penetration depth measurements using tunnel-diode resonator (TDR)}

The London penetration depth $\lambda(T)$ was measured using a version of an RF frequency-domain tunnel-diode resonator designed specifically for thin film measurements. A detailed description of the TDR technique and its applications can be found in Refs.~\cite{Prozorov2000a,Prozorov2006,Prozorov2011,Prozorov2021}. In brief, a thin film sample is inserted between a pair of Helmholtz coils, which are part of an LC tank circuit driven by a tunnel diode biased to its region of negative differential resistance, canceling losses and resulting in stable resonance at approximately 10\,MHz. Changes in the sample's magnetic susceptibility alter the total magnetic inductance, thereby shifting the resonant frequency. With the small amplitude of the excitation field ($\sim 20\,\text{mOe}$ in our case), the film remains in the Meissner state, supported by the absence of any nonlinearities and hysteresis while warming above $T_c$ and cooling back to low temperatures, with or without an AC field present. The frequency shift is proportional to magnetic susceptibility $\chi(T)$ via a geometric factor $G$, $\Delta f(T)/f{0}= G\,\chi(T)$, where $f_{0}$ is the resonant frequency without the sample. Magnetic susceptibility can be converted to an effective penetration depth $\lambda(T)$, which is used to estimate the normalized superfluid density \citep{Prozorov2018}, $\rho_s(T) = \left(\lambda(0)/\lambda(T)\right)^{2}$, where $\lambda(0)$ is the London penetration depth at $T=0$.

\section{Results and discussion}

\subsection{Magnetic flux penetration and trapping}

\begin{figure}[tbh]
\centering
\includegraphics[width=0.95\linewidth]{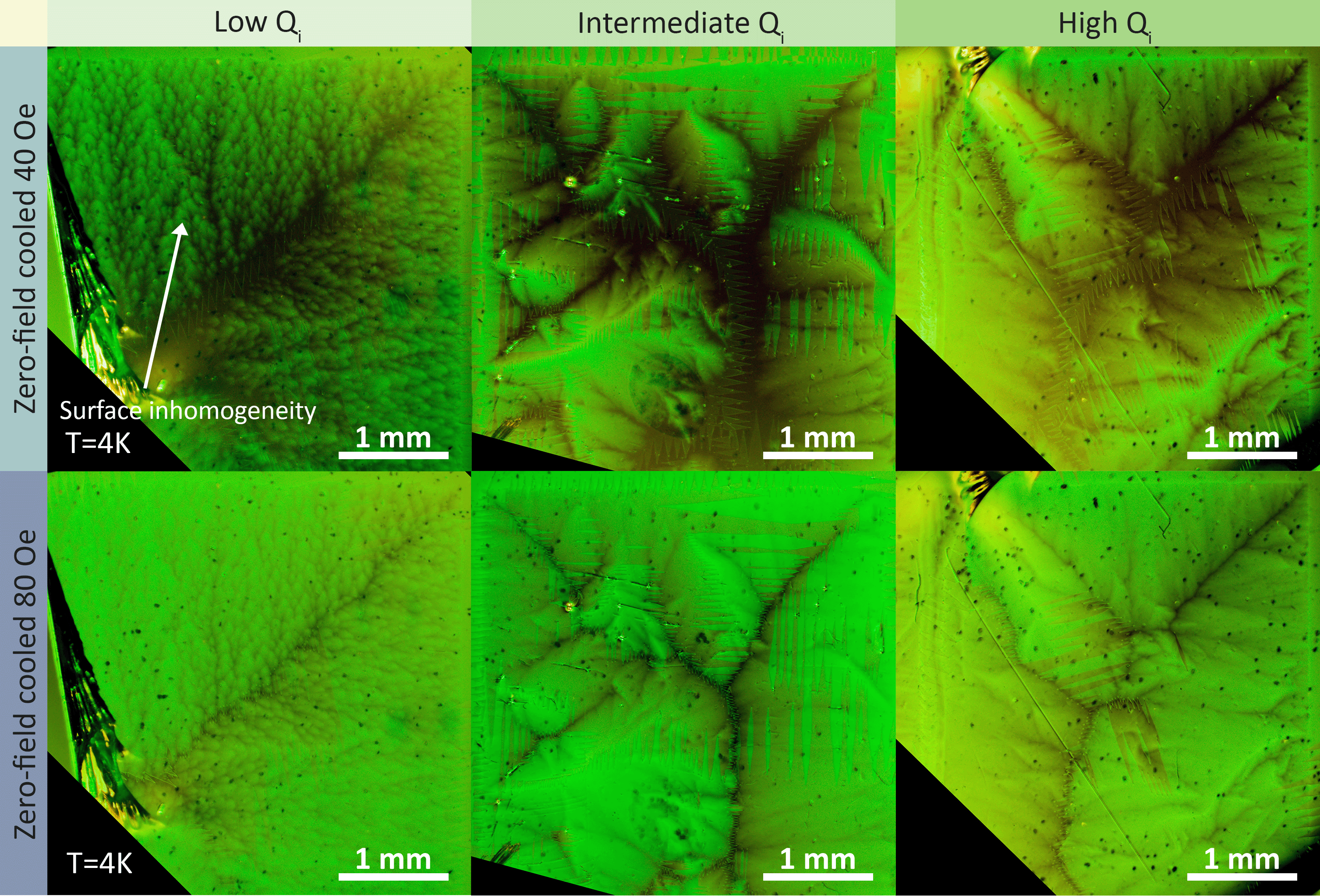}
\caption{\label{fig:MO_ZFC}Magneto-optical imaging of penetrating magnetic flux after cooling in zero-field (ZFC) to 4\,K. Brighter contrast corresponds to higher $B_{z}$. The columns show three types of films, and the rows show flux penetration after applying 40\,Oe (top) and then increasing it to 80\,Oe (bottom). The low-$Q_{\mathrm {i}}$ film is clearly distinct, exhibiting a granular structure, whereas the high-$Q_{\mathrm {i}}$ film displays the lowest flux penetration; see the text for discussion.}
\end{figure}

Figure~\ref{fig:MO_ZFC} shows representative ZFC MO images taken at 4 K for the three films (columns). First, a magnetic field of 40 Oe was applied (row 1 in Figure~\ref{fig:MO_ZFC}), and then it increased to 80 Oe (row 2). The low-$Q_{\mathrm{i}}$ film exhibits a unique morphology, distinct from that of the other two films (and many other Nb films that we have imaged). It shows a distinct granular structure of the magnetic induction. However, the film morphology is atomically uniform, suggesting the clustering of pinning centers or other intrinsic defects. 
%\sout{likely due to increased surface tension during deposition, which results in the formation of droplets of sputtered metal on the surface.} 
Not surprisingly, this film showed the most irregular flux front. The other two films show more regular behavior. The leaf-like structures are caused by flux penetration from larger defects or imperfections. However, it does not affect our measurements of the local flux gradient, which is a measure of the pinning strength.

\begin{figure}[tb]
\includegraphics[width=0.9\linewidth]{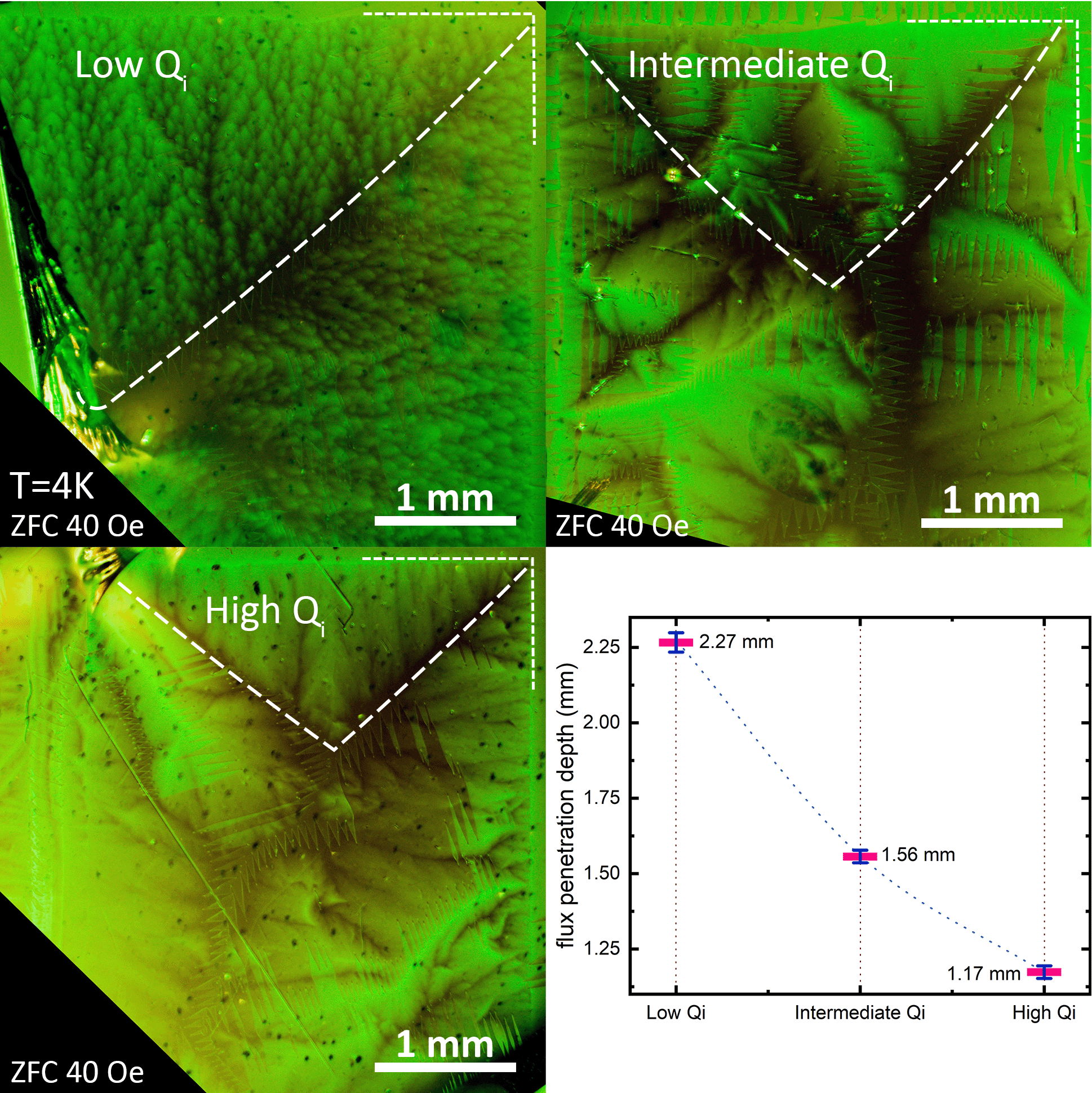}
\caption{\label{fig:flux_depth}Flux penetration depth as a function of resonator quality factor. The depth of the flux front decreases with increasing $Q_i$, indicating that high-$Q$ films possess superior screening capacity and higher critical current densities.}
\end{figure}

To analyze the MO images, we estimate the penetration depth of the magnetic flux inside the sample (flux front) at 40 Oe, which is low enough so that the flux does not overlap in the center, as shown in Fig.~\ref{fig:flux_depth}. Not surprisingly, the low-$Q_{\mathrm i}$ film exhibits the largest, i.e., the deepest flux penetration compared to the other two films. This is quantified in the lower right panel, which shows the estimated depth of flux penetration, as depicted in the images. There is a correlation between this depth and $Q_{\mathrm i}$. The shorter depth of flux penetration implies a larger vortex gradient; hence, the critical (persistent) current density.

The critical (persistent) current density depends on two opposing factors: a higher defect concentration (a dirtier regime) and a larger condensation energy (stronger superconductivity). For qubit applications, the latter is definitely a positive factor. The former is not quite definitive. On the one hand, from the perspective of qubit circuits, defects increase the probability of trapping vortices \cite{wang2014}. Because vortex motion produces microwave loss and excess noise, even a small number of mobile vortices can strongly affect $Q_{\mathrm i}$ \cite{Stan2004,Bothner2011,bahrami2025,oh2024}. However, which mechanism prevails is not obvious and requires thorough characterization using multiple techniques. In the present case, it appears that the condensation energy gain outweighs any potential differences in the defect structure.

\begin{figure}[tb]
\centering
\includegraphics[width=0.9\columnwidth]{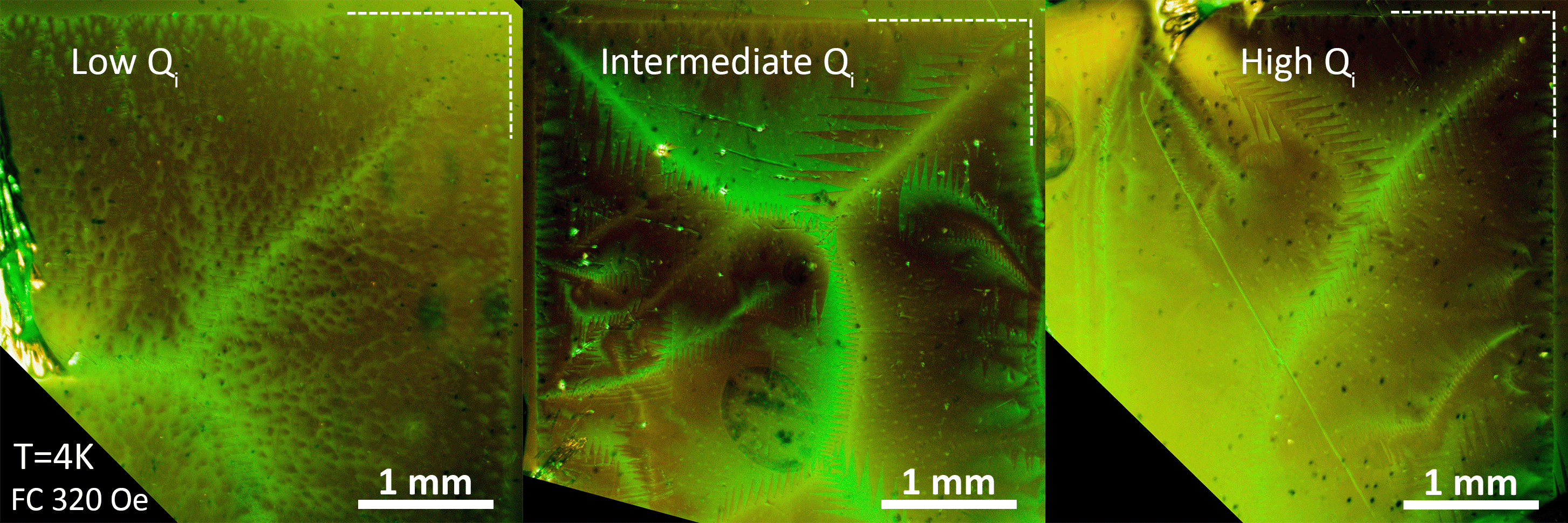}
\caption{\label{fig:MO_FC}Magneto-optical Faraday images of the remanent state after cooling in a 320 Oe to 4\,K and turning the field off. The low-quality factor film shows prominent granular structure and mesoscopic inhomogeneity. However, at the macro scale, all films exhibit a roof-like shape of trapped flux, as expected from the Bean model. }
\end{figure}

Finally, for completeness, Fig.~\ref{fig:MO_FC} shows the MO images after cooling in a magnetic field of $H=320\,\text{Oe}$ at 4\,K and then turning it off (field-cooling process). The low-$Q_{\mathrm i}$ film shows a granular structure, similar to that in the ZFC images, Fig.~\ref{fig:MO_ZFC}. The ``grains" appear dark, indicating a significantly reduced ability to trap magnetic flux. They would be brightest if they were superconducting. The images of the other two films are also consistent with the ZFC results.  

\subsection{London penetration depth}

\begin{figure}[tb]
\centering
\includegraphics[width=0.9\columnwidth]{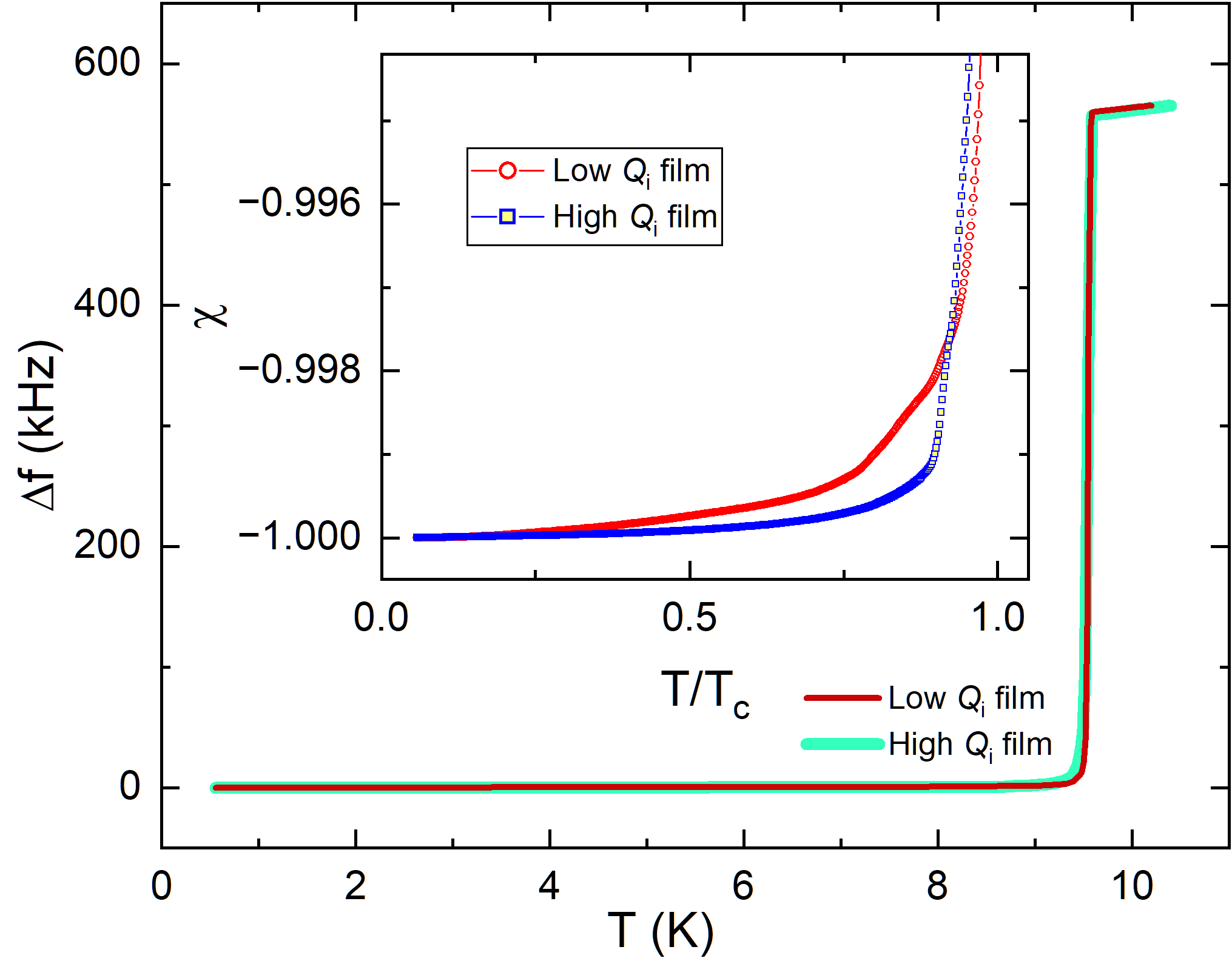}
\caption{\label{fig:chi}Main frame: raw data - resonant frequency shift as a function of temperature. Inset: low-temperature variation of the magnetic susceptibility at low temperatures. Note the extremely small range shown, from -1.000 to -0.994.}
\end{figure}

To further understand the physics behind the differences in the studied films, we will now discuss quasiparticle spectroscopy performed by the TDR susceptometer. The main frame of Fig.~\ref{fig:chi} shows the temperature-dependent resonant frequency shift. All samples exhibit a sharp superconducting transition around $ T_c\simeq 9.4$~K and, at first sight, do not show additional features. However, the inset, which zooms in below the transition temperature, shows irregular features and a broadening of $\chi(T)$ that deviates significantly from the BCS $s-$wave curve. Note a small range of magnetic susceptibility in the zoomed figure, from -1.000 to -0.994; such variation is undetectable in conventional magnetometry/susceptometry.

The irregular behavior of $\chi(T)\sim \lambda (T)$ (a convex downturn near transition) implies irregular London penetration depth, $\lambda (T)$, which in turn signifies the irregular temperature dependence of the superfluid density, $\rho (T)=(\lambda(0)/\lambda (T))^2$. The superfluid density, normalized by its value in the clean case, is:
\begin{equation}
n_{s}=1+\intop_{-\infty}^{\infty}\frac{\partial f\left(E\right)}{\partial E}\frac{N\left(E\right)}{N_{n}}\,dE
\label{eq:ns}
\end{equation}
\noindent where $E=0$ corresponds to the Fermi level, $N(E)$ is the quasiparticle density of states, $N_n$ is its normal-state value, and $f(E)$ is the Fermi function. The temperature dependence enters primarily through the derivative of the Fermi function, $\partial f(E)/\partial E = -\mathrm{sech}^{2}(E/2t)/(4t)$. At $t \to 0$, the derivative approaches a delta function. At finite temperatures, it broadens into a bell-shaped kernel that effectively samples the density of states starting from $E_F$, revealing any significant subgap features. This is why we refer to this approach as ``gap spectroscopy.'' 
Considering sharp superconducting transitions and nearly complete screening of the magnetic field in all films, the subtle irregularity cannot be due to significant physical-chemical variations of the films' atomic structure \cite{ghimire2021}. Instead, as discussed in Ref.~\citep{ghimire2024quasi}, this behavior is consistent with the presence of additional quasiparticle states inside the gap.

\subsection{Implications for Nb-based qubits loss}

The combined results of MO and TDR constrain plausible microscopic explanations for variations in $Q_{\mathrm i}$ of the films studied. The MO data directly connect low $Q_{\mathrm i}$ to the weakest ability to screen the magnetic field and enhance spatial heterogeneity on a larger mesoscopic scale. 

The nonmonotonic behavior of the superfluid density is often associated with disorder-driven pair breaking, proximity-coupled surface layers, or oxide/suboxide-related states that increase the quasiparticle density at low temperatures. 
A natural candidate for the nonmonotonic behavior of the superfluid density is the same defect landscape that limits $Q_i$ in microwave resonators: the two-level systems (TLS) and the structurally and chemically disordered layers that host them. Recent studies have provided direct evidence that Nb/oxide interfaces can host interacting TLS-like defect ensembles \cite{Gorgichuk2023}, with oxygen vacancies implicated as a prominent microscopic source of loss \cite{Bafia2024}. 

Although TLS are most commonly discussed as a dielectric loss mechanism, other defects can influence the superconducting quasiparticle spectrum measured by a macroscopic probe, such as ours. Local disorder in the surface oxide can modify chemical bonding and the effective pairing environment, introduce resonant scattering channels, and potentially stabilize localized magnetic moments. Any of these effects can produce subgap states and broaden the coherence peaks, leading to an increased quasiparticle population that manifests as a measurable, non-activated, and sometimes non-monotonic response in $\lambda(T)$ at low temperatures. Further discussion of such defects can be found in Refs.~\citep{Shiba1968, Rusinov1969, Yu1965, zarea2025}.

\section{Conclusions}

We have correlated mesoscopic and macroscopic vortex behavior with low-energy electrodynamics and microwave quality factors in epitaxial Nb films grown at different deposition temperatures. Magneto-optical imaging revealed significant differences between the film deposited at the highest temperature, 730\,$^o$C, and the other two films, deposited at 630\,$^o$C, and 520\,$^o$C, respectively. There is significant inhomogeneity and a droplet-like structure in the first film. The difference between the latter two films is less pronounced, but the film with the highest-$Q_{\mathrm i}$ shows better screening and stronger pinning. This supports the conclusions of several recent papers on Nb and Ta films that pinning might be beneficial for the quantum performance of qubits, presumably by immobilizing Abrikosov vortices \citep{Abrikosov2017, AbrikosovGorkov1960ZETF,Murthy2025, bahrami2025}. 

In lower-$Q_{\mathrm i}$ films, London penetration depth spectroscopy reveals the presence of additional localized states in the energy gap, likely due to two-level systems and/or Shiba-like pair-breaking by magnetic moments. Their microscopic origin requires further studies.

This work highlights the usefulness of our recently developed method for studying planar superconducting devices \cite{Joshi2023, ghimire2024quasi, Datta2024}: a combination of magneto-optical imaging and precision quasiparticle spectroscopy, which provides an effective characterization and screening approach for optimizing the fabrication protocols of superconducting quantum circuits.

\begin{acknowledgments}
This work was supported primarily by the U.S. Department of Energy, Office of Science, National Quantum Information Science Research Centers, Superconducting Quantum Materials and Systems Center (SQMS), under Contract No. 89243024CSC000002. Ames Laboratory is supported by the U.S. Department of Energy (DOE), Office of Science, Basic Energy Sciences (BES), Materials Science \& Engineering Division (MSED) and is operated by Iowa State University for the U.S. DOE under contract DE-AC02-07CH11358. M.A.T. and K.R.J. were supported by DOE, BES, MSED at Ames National Laboratory.  
\end{acknowledgments}

%\bibliographystyle{apsrev4-2}
%\bibliography{Nb-NIST}


\end{document}